\documentclass[doublecol]{epl2}
\usepackage{graphicx}
\usepackage{amsmath}    
\usepackage{amssymb}    
\usepackage{verbatim}   
\usepackage{color}      
\usepackage{subfigure}  
\usepackage{amsfonts}
\newcommand{\vct}[1]{{\bf #1}}
\usepackage{subfigure}

\renewcommand\Im{\operatorname{Im}}

\title{Small distance expansion for radiative heat transfer between curved objects}

\author{Vladyslav A. Golyk\inst{1} \and Matthias Kr\"uger\inst{1} \and Alexander P. McCauley\inst{1} \and Mehran Kardar\inst{1}}
\shortauthor{V. A. Golyk, M. Kr\"uger, A. P. McCauley and M. Kardar}

\institute{ 
  \inst{1} Massachusetts Institute of Technology, Department of
  Physics, Cambridge, Massachusetts 02139, USA
}

\abstract{
We develop a small distance expansion for the radiative heat transfer between gently curved objects, in terms of the ratio of distance to radius of curvature. A gradient expansion allows us to go beyond the lowest order proximity transfer approximation. The range of validity of such expansion depends on temperature as well as material properties. Generally, the expansion converges faster for the derivative of the transfer than for the transfer itself, which we use by introducing a {\em near-field adjusted plot}. For the case of a sphere and a plate, the logarithmic correction to the leading term has a very small prefactor for all materials investigated.}


\begin{document} 

\maketitle
More than 40 years ago Van Hove and Polder used Rytov's fluctuational electrodynamics\cite{Rytov} to predict that the radiative heat transfer (HT) between objects separated by a vacuum gap can exceed the blackbody limit\cite{VanHove}. This is due to evanescent electromagnetic fields decaying exponentially into the vacuum. HT due to evanescent waves has also attracted a lot of interest due to its connection with scanning tunnelling microscopy, and scanning thermal microscopy under ultra-high vacuum conditions\cite{Pendry, Majumdar}.  The enhancement of HT in the near-field regime (generally denoting separations small compared to the thermal wavelength, which is roughly 8 microns at room temperature) has only recently been verified experimentally\cite{exp1,exp2
}. Theoretically,  HT has been considered for a limited number of shapes: Parallel plates\cite{VanHove, Loomis94, Pendry,BiehsRough}, a dipole or sphere in front a plate \cite{partplate1,scattering2,Fan}, two  dipoles or spheres  \cite{partplate1,twospheres2,twospheres3}, and for a cone and a plate \cite{alexcode}. The scattering formalism has been successfully exploited\cite{scattering1,scattering3,scattering2,Reynaud,longpaper} in this context. Although powerful numerical techniques \cite{alexcode,Johnson} exist for arbitrary geometries, analytical computations are limited to planar, cylindrical and spherical cases \cite{longpaper}. Alternatively, the HT between closely spaced curved objects can be computed by use of the proximity transfer approximation (PTA) \cite{scattering2,PTA1,Fan}, which exploits an approach that has been extensively used in the context of fluctuation induced forces \cite{PFA} (referred to as proximity force approximation): The HT between two parallel plates (per unit area), $H_{pp}(S)$, for separation $S$ is averaged over one of the (projected) curved surfaces. PTA is generally assumed to hold asymptotically for small separations, however no rigorous derivation appears available in the literature.

Here we develop a gradient expansion for HT between closely spaced objects, which enables to rigorously justify PTA and to quantify corrections to it in the limit of small separation. This method, which has been proposed for Casimir forces\cite{Argentina,BimGrad2,BimGrad1}, exploits the mapping of coefficients of a perturbative expansion on one side and a gradient expansion on the other. We carefully explore the limitations and subtleties of this method in the case of HT and propose a {\em near-field adjusted plot} that overcomes the possibly slow convergence of the expansion.


Consider two non-magnetic objects with dielectric permittivities $\epsilon_1(\omega)$ and $\epsilon_2(\omega)$ at temperatures  $T_1$ and $T_2$, respectively.  The HT {$Q$} between them is symmetric and can be written as an integral over frequency $\omega$, as
\begin{equation}\label{bimonte45}
{Q}=\int_{0}^{\infty}d\omega\left[n_\omega(T_1)-n_\omega(T_2)\right]{q},
\end{equation}
where $n_\omega(T)= \left(e^{\hbar\omega/k_BT}-1\right)^{-1}$ is the Bose-Einstein weight. The spectral density $q$ depends only on the classical scattering amplitudes of the objects and their relative positions \cite{longpaper}. For the sake of simplicity, let  object 1 be a semi-infinite, planar body, filling the space $z<0$, whereas object 2 has a curved surface with the smooth height profile $z=S(\textbf{x})$; $\textbf{x}=(x,y)$ are Cartesian coordinates in the plane normal to the $z$ axis.
PTA is then calculated as
\begin{equation}\label{PTAdef}
Q_{PTA}=\int_{\Sigma} d^2\textbf{x}~ H_{pp}(S),
\end{equation}
and analogously for the spectral density, where $\Sigma$ is the projected surface.

\begin{figure}\centering
\includegraphics[width=8 cm]{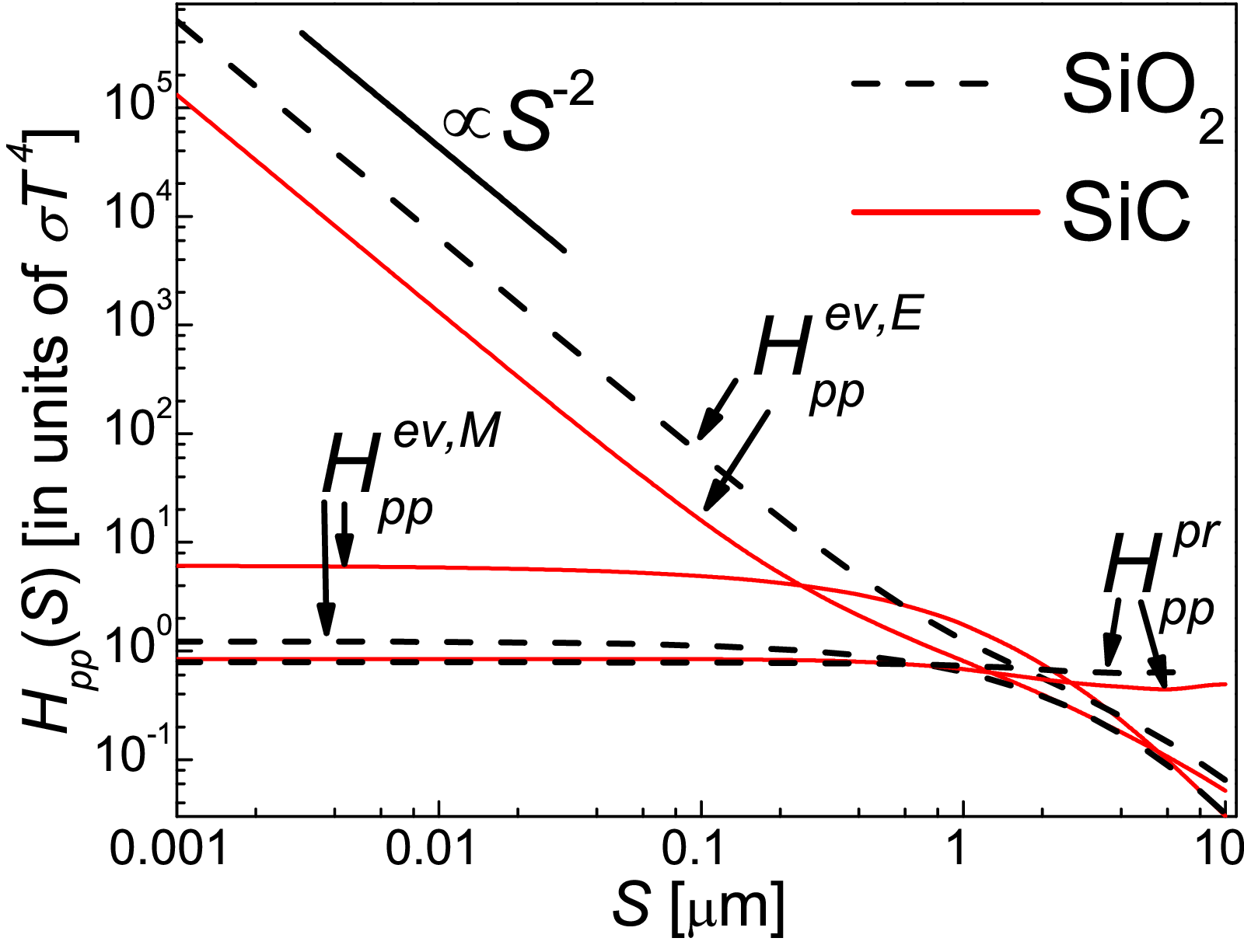}
\caption{The evanescent electric $\left(H_{pp}^{ev,E}\right)$, magnetic $\left(H_{pp}^{ev,M}\right)$ and total propagating $\left(H_{pp}^{pr}\right)$ contributions to heat transfer between parallel plates per unit area, $H_{pp}=H_{pp}^{ev,E}+H_{pp}^{ev,M}+H_{pp}^{pr}$, as a function of separations for $T_1=0$K and $T_2=300$K. Data are shown for two materials, SiC and SiO$_2$, and are normalized by the result for black bodies (where $\sigma=\pi^2 k_B^4 / (60 \hbar^3 c^2)$).}\label{fig0}
\end{figure}

Different contributions \cite{partplate1} to $H_{pp}(S)$ are separately depicted in Fig.~\ref{fig0} for two dielectric materials, SiC \cite{SiCdiel} and SiO$_2$ (computed using optical data\footnote{The optical data was provided by P. Sambegoro from Prof. Gang Chen's Nanoengineering Group and had been previously used in Refs.\cite{exp1,scattering2,longpaper}.}). Propagating waves give rise to an almost separation independent contribution, whereas the transfer due to evanescent waves depends strongly on separation. Both parts have in turn two distinct components from electric ($E$) and magnetic ($M$) modes. {Note that the modes called $E$ and $M$ here, are in the literature often referred to as transverse magnetic (TM) and transverse electric (TE), respectively.} Notably, if $S$ is the smallest scale, the evanescent $E$-mode diverges as $S^{-2}$, while all other contributions approach finite values. For the two materials considered, the figure indicates that the $S^{-2}$ contribution is dominant for $S\lesssim 0.03{~\mu}${m} and $S\lesssim0.3{~\mu}${m}, respectively. In general, the range of this regime, sometimes referred to as the nonretarded regime, depends strongly on material properties \footnote{An expansion of the Fresnel coefficients for large evanescent wavevector shows that the (spectral) $S^{-2}$ contribution is dominant if (for two equal materials) $\Im[({\epsilon}-1)/({\epsilon}+1)] \gg \Im[{\epsilon}-1] (\omega d/2c)^2$ holds.}.  Because the behavior of $H_{pp}(S)$ is nontrivial in general, we restrict ourselves to the case $H_{pp}(S)\propto S^{-2}$, such that Eqs.~\eqref{ansatz}-\eqref{eq:cyl} below are valid. At the end, we  give an outlook to larger separations. The expansion can  equally well be performed for the transfer in Eq.~\eqref{bimonte45} as for its spectral density, and we choose the latter due to its larger information content. The spectral density for two parallel plates reads accordingly
\begin{equation}\label{scaling}
h_{pp}(S)=\frac{\lambda_\omega}{S^2},
\end{equation}
where the coefficient $\lambda_\omega$ uniquely describes the material dependence of the HT in the considered range.
Following Refs.~\cite{Argentina,BimGrad2,BimGrad1}, we assume (justified a posteriori) that $q$ {in Eq.\eqref{bimonte45}} admits a local expansion in $\nabla S$,
\begin{equation}\label{ansatz}
\begin{split}
q[S{(\textbf{x})}]&=q_{PTA}[S{(\textbf{x})}]\\&+\int_{\Sigma}d^2 \textbf{x}~ \beta_\omega~ h_{pp}(S)~\nabla S\cdot\nabla S+\dots,
\end{split}
\end{equation}
where $\beta_\omega$ is a \textit{geometry-independent} expansion parameter and the dots represent higher order terms in $\nabla S$. We will see below that Eq.~\eqref{ansatz} is an expansion in the ratio of separation to radius of curvature of the surface.


The coefficient $\beta_\omega(S)$ is obtained from the {\it perturbative} expansion \cite{BiehsRough} of $q[S]$ for $|s(\textbf{x})|\ll d$, with $s$ defined with respect to the point of closest proximity $d$, $S(\textbf{x})=d+s(\textbf{x})$: If both a perturbative expansion as well as the gradient expansion in Eq.~\eqref{ansatz} exist, the two expansions must have identical coefficients in their common limit, i.e. the expansion of $q$ in Eq.~\eqref{ansatz} has to match the perturbative expansion at small momenta $\vct{k}$, which is given by
\begin{equation}\label{Gkd}
\begin{split}
q[d+s(\textbf{x})]&=\alpha_0(d)+\alpha_1(d)\tilde{s}(\textbf{0})\\&+\int {\frac{d^2 \textbf{k}}{(2\pi)^2}}~\alpha_2(k;d)~|\tilde{s}(\textbf{k})|^2
,
\end{split}
\end{equation}
where, e.g., {$\alpha_0(d)$ is proportional to $h_{pp}(d)$ times the corresponding projected area, and} $\tilde{s}(\textbf{k})$ is the Fourier transform of $s(\textbf{x})$ with in-plane wave-vector $\textbf{k}$. 
Thus, if it is possible to expand  $\alpha_2(k;d)$ to second order in $k$, we can match the expansion
\begin{equation}\label{expansion}
\alpha_2(k;d)=\alpha_2^{(0)}(d)+\alpha_2^{(2)}(d)k^2+\dots,
\end{equation}
to Eq.~\eqref{ansatz} to get
\begin{equation}\label{alphabeta}
\beta_\omega h_{pp}(d)={\alpha_2^{(2)}(d)}.
\end{equation}
{The gradient expansion of Eq.~\eqref{ansatz} requires knowledge of the coefficient $\beta_\omega$, which, by Eq.~\eqref{alphabeta}, is connected to $\alpha_2^{(2)}(d)$. $\alpha_2^{(2)}(d)$ will in turn be computed below. The coefficients $\alpha_0$ and $\alpha_1$ in Eq.~\eqref{Gkd} are hence irrelevant for the remainder of the computation, and we omit their detailed analysis.} 

In the asymptotically close regime, $\beta_\omega$ is $d$-independent, as both $h_{pp}(d)$ and $\alpha_2^{(2)}(d)$ are proportional to $d^{-2}$.
The perturbative calculation for $q[S]$  is done by use of a plane-wave basis with in-plane wave-vector $\textbf{k}$ (eventually identical to $\textbf{k}$ in Eq.~\eqref{Gkd}) and modes $p= E,M$. The reflection matrix of the planar object 1 is diagonal in this basis and given by the Fresnel coefficients $r^{(1)}_{p}(\textbf{k})$. The corresponding coefficient $R^{(2)}_{pp'}(\textbf{k},\textbf{k}')$ for the curved object 2 depends on both incoming (primed) and scattered waves, and can be expanded in powers of $\tilde{s}$\cite{voronovich} (with $k\equiv|\textbf{k}|$), as
\begin{equation}\label{voron}
\begin{split}
&R^{(2)}_{pp'}(\textbf{k},\textbf{k}')=(2\pi)^2\delta^{(2)}(\textbf{k}-\textbf{k}')\delta_{pp'}r^{(2)}_{p}(k)\\
&+B_{pp'}^{(i)}(\textbf{k},\textbf{k}')\tilde{s}(\textbf{k}-\textbf{k}')\\
&+\int\frac{d^2\textbf{k}''}{(2\pi)^2}B_{pp'}^{(ii)}(\textbf{k},\textbf{k}';\textbf{k}'')\tilde{s}(\textbf{k}-\textbf{k}'')\tilde{s}(\textbf{k}''-\textbf{k}')+\cdots.\\
\end{split}
\end{equation}
The expressions for $B_{pp'}^{(i)}$ and $B_{pp'}^{(ii)}$ are given in Ref.~\cite{voronovich}.
Substituting Eq.~\eqref{voron} into the HT-expression for corrugated surfaces\cite{scattering1} and expanding to second order in $\tilde{s}(\textbf{k})$ yields  $\alpha_2(k;d)$. Expanding the latter for small $k$ yields $\alpha_2^{(2)}(d)$ according to Eq.~\eqref{expansion}, and with Eq.~\eqref{alphabeta} the desired $\beta_\omega$ is determined. 

For the experimentally most relevant case of a sphere of radius $R$ in front of a plate, Eq.~\eqref{ansatz} is evaluated to yield the following expansion in $d/R$ (where $S(\textbf{x})=d+R(1-\sqrt{1-|\textbf{x}|^2/R^2})$),
\begin{equation}\label{fulratio}
\begin{split}
q(d)=&\frac{2\pi R \lambda_\omega}
{d}\left[1-(2\beta_\omega-1)\frac{d}{R}\log \frac{d}{d_0}\right]+\mathcal{O}(d^0).
\end{split}
\end{equation}
Here $d_0$ is an unknown constant of integration. PTA in Eq.~\eqref{PTAdef} corresponds to the result for $\beta_\omega=0$ (compare to Eq.~\eqref{ansatz}).
Hence in Eq.~\eqref{fulratio} $2\pi\lambda_\omega\log \left(d/d_0\right)$ represents a trivial correction to the leading $2\pi R\lambda_\omega/d$ and appears due to the projection of the sphere's surface onto the $xy$-plane.\addtocounter{footnote}{-1}

As the term of $\mathcal{O}(d^0)$ in Eq.~\eqref{fulratio} is unknown from the expansion and  might not be small compared to the other terms in Eq.~\eqref{fulratio} (we compare it to a logarithmic term), possibly leading to slow convergence in practical cases, a better quantity to consider is its derivative,
\begin{equation}\label{derratio}
q^{\prime}(d)=-\frac{2\pi R \lambda_\omega}
{d^2}\left[1+(2\beta_\omega-1)\frac{d}{R}\right]+\dots,
\end{equation}
which converges faster, as the constant term in Eq.~\eqref{fulratio} drops out.


\begin{figure}\centering
\includegraphics[width=8 cm]{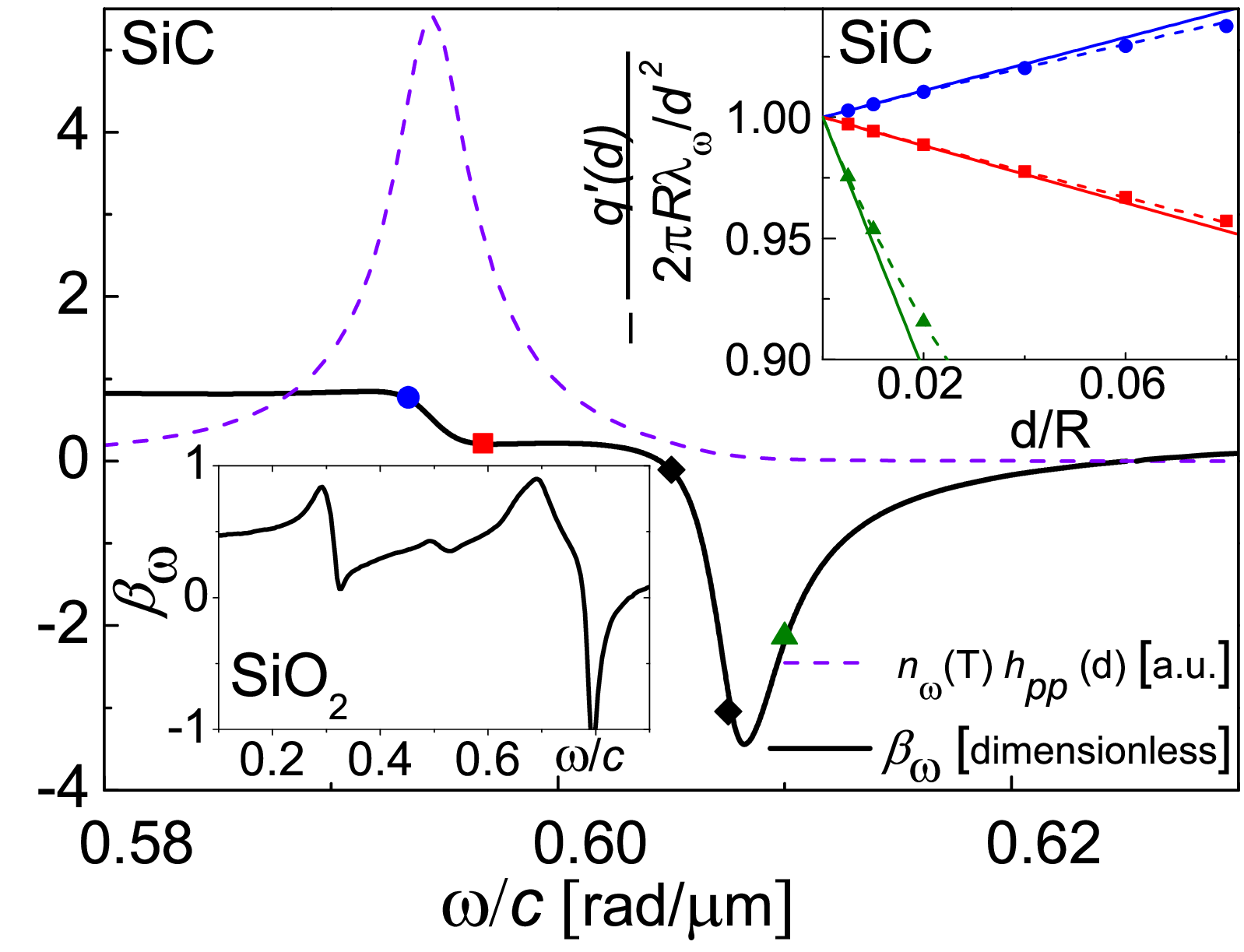}
\caption{(color online). The gradient expansion parameter $\beta_\omega$ as a function of frequency for SiC, computed via Eq.~\eqref{alphabeta}, (solid line), and as found by fitting exact data for a sphere in front of a plate to Eq.~\eqref{derratio} (data points). The dashed line shows $h_{pp}(d=10{~\mathrm{nm}})n_\omega(T=300K)$, illustrating the dominant range for near-field transfer of SiC. The upper inset provides three examples of fitting the exact data (data points) to  Eq.~\eqref{derratio}, where solid (dashed) lines omit (include) the assumed next order term in Eq.~\eqref{derratio}\protect\footnotemark. The point styles used are in accordance with the corresponding frequencies in the main figure. In the considered regime (Eq.~\eqref{scaling}) these curves depend on $R$ only through $d/R$. For $\omega{/c}=\{0.5934,0.5967,0.6050,0.6075,0.6100\}\mathrm{rad}/ \mu$m shown in the main plot, the values computed in the gradient expansion ($\beta_\omega=\{0.774,0.206,-0.086,-3.026,-2.156\}$) agree well with the fitted ones ($\{0.773,0.206,-0.105,-3.044,-2.112\}$). The lower inset shows $\beta_\omega$ for SiO$_2$.}\label{beta}
\end{figure}

{In order to test these results, we semi-analytically compute the exact heat transfer between a sphere and a plate. This scheme, where the scattering properties of sphere and plate are expanded in cylindrical multipoles \cite{alexcode}, directly yields the complete heat transfer between the two objects as a sum over multipole-contributions (in contrast to other methods which e.g. use surface discretization). Here,} a maximal {multipole} order of $l_{\rm max}=2500$ is used for the smallest separations $d/R=0.004$ shown to obtain precise data (including the derivative). We can hence validate Eq.~\eqref{derratio} (or Eq.~\eqref{fulratio}) by finding $\beta_\omega$ in two independent ways. The main panel of Fig.~\ref{beta} shows $\beta_\omega$ computed by the gradient expansion for SiC, together with five values obtained by fitting of exact data to Eq.~\eqref{derratio}. We observe good agreement\addtocounter{footnote}{-1}\footnotemark\footnotetext{For improved fits we assume a next correction to Eq.~\eqref{derratio} of the form $\gamma\log\left(d/R\right)$ with adjustable $\gamma$.}. The plotted frequency region is the relevant interval for SiC, with one dominant resonance around $\omega{/c}=0.6$~rad/$\mu$m, as indicated by $h_{pp}(d=10{~\mathrm{nm}})$ multiplied by the weight $n_\omega(T=300K)$ (dashed curve). The upper inset shows the result of Eq.~\eqref{derratio} using the fitted $\beta_\omega$ together with the exact data for three frequencies also depicted in the main plot. We have thus demonstrated the validity of the gradient expansion both qualitatively and quantitatively. The lower inset shows $\beta_\omega$ for SiO$_2$ computed by the gradient expansion.

The total transfer $Q$ follows from  Eqs.~\eqref{bimonte45} and \eqref{fulratio}, where we still consider $H_{pp}(S)\propto S^{-2}$. For materials with $|{\epsilon}|$ not too large (e.g. SiO$_2$), this regime requires $S\ll \{\lambda_{T_1},\lambda_{T_2}\}$ ($\lambda_T=\hbar c /(k_BT)$ is the thermal wavelength, $\lambda_T\approx7.6{~\mu}$m at $300K$. It sets the range of wavelengths contributing to the transfer in Eq.~\eqref{bimonte45}). Integrating Eq.\eqref{fulratio} over frequencies $\omega$, we obtain
\begin{equation}\label{fulratio2}
\begin{split}
Q(d)=\frac{2\pi R \lambda}{d}\left[1-(2\beta-1)\frac{d}{R}\log \frac{d}{d_0}\right]+\mathcal{O}(d^0),
\end{split}
\end{equation}
where $\lambda=\int_{0}^{\infty}d\omega\left[n_\omega(T_1)-n_\omega(T_2)\right]
\lambda_\omega$ and 
$\beta=\lambda^{-1}\int_{0}^{\infty}d\omega\left[n_\omega(T_1)-n_\omega(T_2)\right]\beta_\omega
\lambda_\omega$.

\begin{figure}\centering
\includegraphics[width=8 cm]{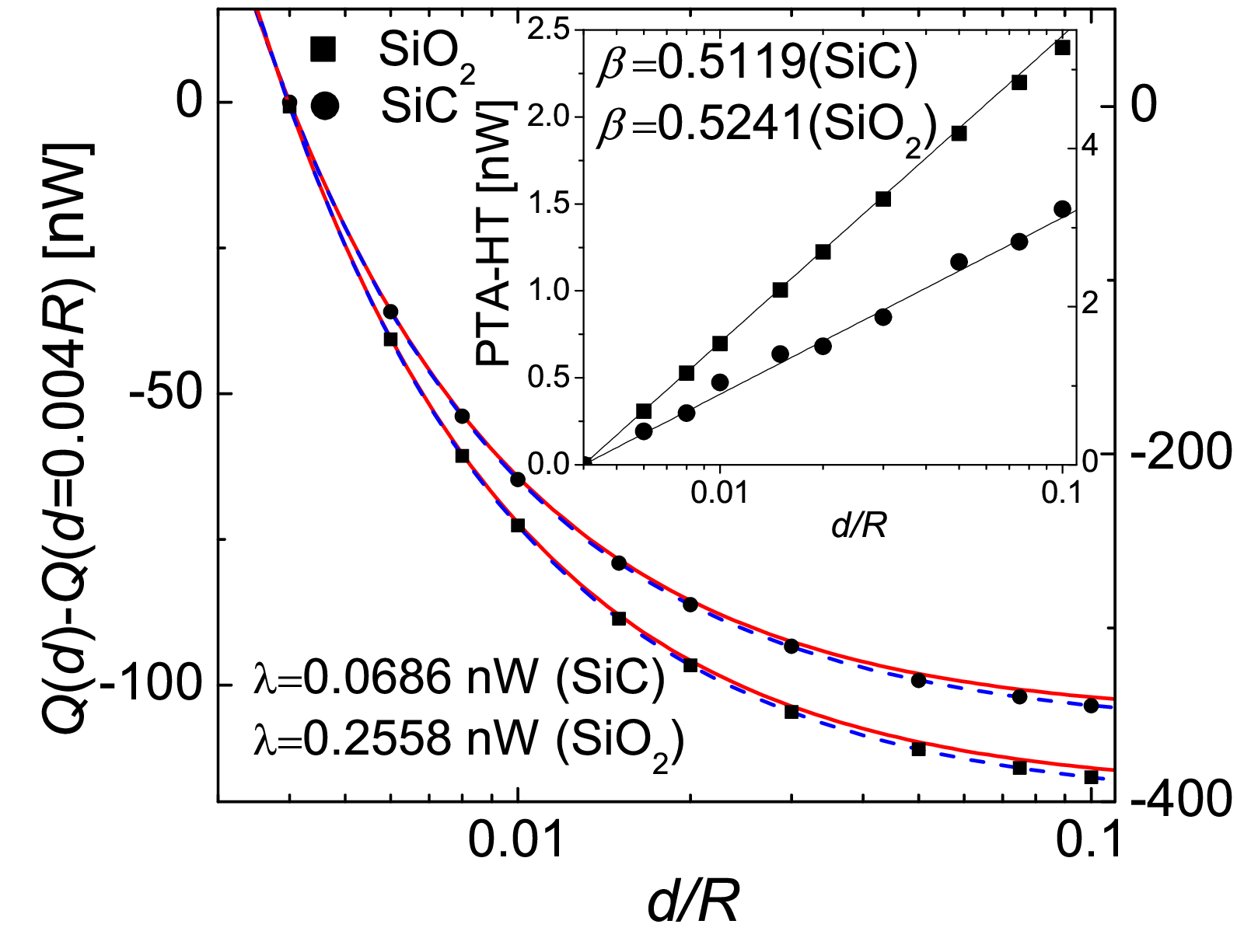}
\caption{Near-field adjusted plot ($Q(d)-Q(d=0.004R)$) for a sphere in front of a plate for SiC (circles, left abscissa) and SiO$_2$ (squares, right abscissa), $T_1=300$K and $T_2=0$K. Exact data, PTA and Eq.~\eqref{fulratio2} are shown as data points,  solid and dashed lines, respectively. The inset depicts the deviation of PTA from the exact data, where lines show the analytical form $4\pi\beta\lambda\log[d/0.004R]$ from Eq.~\eqref{fulratio2}. In the regime considered (Eq.\eqref{scaling}), all curves shown depend on $R$ only through $d/R$.}
\label{full}
\end{figure}

Figure \ref{full} depicts the HT from a sphere to a plate as a function of $d/R$ for $T_1=300$K and $T_2=0$. In order to eliminate the constant term in Eq.\eqref{fulratio2}, the main panel depicts a \emph{near-field adjusted plot}, where we show $Q(d)-Q(d=0.004R)$. Such a plot takes advantage of the faster convergence observed for the derivative in Eq.~\eqref{derratio}, and allows to accurately describe the exact data by use of Eq.~\eqref{fulratio2}, and in the same manner, experimental data could be compared to Eq.~\eqref{fulratio2}. We note that the reference separation (here $d=0.004R$) should be chosen in the range where the $d/R$-expansion is valid. For both SiC and SiO$_2$, the exact data points agree remarkably well with Eq.~\eqref{fulratio2}, over the whole range of separations depicted (up to $d/R=0.1$). We also show PTA of Eq.~\eqref{PTAdef}, which -- again -- is to the lowest two orders in $d/R$ given by Eq.~\eqref{fulratio2} with $\beta=0$. Its deviation from the exact data is hence quantified by Eq.~\eqref{fulratio2} to $4\pi\beta\lambda\log[d/0.004R]$. This is shown in the inset, with convincing agreement to the exact data, which is scattered due to numerical precision (the inset probes differences of $\sim1\%$ of the total transfer). Based on Fig.~\ref{fig0}, we expect these curves to describe the HT for separations up to $d\lesssim 0.03{~\mu}$m and $d\lesssim 0.3{~\mu}$m for SiC and SiO$_2$ respectively. E.g., for $R=1{~\mu}$m ($R=10{~\mu}$m) for SiC (SiO$_2$), this corresponds to $d/R\lesssim 0.03$.

The subleading term in Eq.~\eqref{fulratio2} is a logarithmic correction. Surprisingly, this term is hardly noticeable due to the coincidence that $\beta=0.5119$ and $\beta=0.5241$ for SiC and SiO$_2$, respectively (despite very different curves for $\beta_\omega$ for the two materials in Fig.\ref{beta}), and hence the prefactor $(2\beta-1)$ is very small (0.02 and 0.05). In other words, the second order term in Eq.~\eqref{fulratio2} is very small because the term with $\beta$, coming from the second term in Eq.~\eqref{ansatz}, almost cancels the one coming from the first term in Eq.~\eqref{ansatz}. Due to this coincidence, which is special for the sphere-plate geometry, PTA predicts a logarithmic term which in reality is almost absent.

Other geometries can readily be analyzed with Eq.~\eqref{ansatz}. For example, for two spheres of radii $R_1$ and $R_2$, we find,
\begin{equation}
\begin{split}
&Q(d)=\frac{2\pi\lambda}{d}  \frac{R_1 R_2}{R_1+R_2}\left[1+\frac{d}{R_1+R_2}\log\frac{d}{d_0}\right.\\&\left.-\left(2\beta-1\right)\left(\frac{d}{R_1}+\frac{d}{R_2}\right)\log\frac{d}{d_0}\right]+\mathcal{O}(d^0),
\end{split}
\end{equation}
where, in contrast to the sphere-plate case, the logarithmic term is pronounced for $\beta\approx0.5$, as has been observed in numerical data \cite{PTA1}. Furthermore, for a cylinder of radius $R$ and length $L$ in front of a plate, the expansion convergences better than Eq.~\eqref{fulratio2},  because the second term has no logarithm,
\begin{equation}
\frac{Q(d)}{L}=\frac{\pi\sqrt{R}\lambda}{\sqrt{2}d^{3/2}}\left[1+\left(2\beta-\frac{3}{4}\right)\frac{d}{R}\right]+\mathcal{O}(d^0),\label{eq:cyl}
\end{equation}
Geometries for which semi-analytical methods of scattering theory are not possible (e.g. paraboloids or ellipsoids), can also be studied with the aid of Eq.~\eqref{ansatz}, and will be left for future work.

For larger separations, the $d$-dependence of the expansion parameter $\beta_\omega$ in Eq.~\eqref{alphabeta} has to be taken into account, and Eq.~\eqref{ansatz} has to be evaluated numerically. This analysis, which poses a number of additional challenges\footnote{Preliminary results indicate an unphysical divergence in the computation of $\beta(d)$ outside the regime where Eq.~\eqref{scaling} holds.} (as well as  insights), is left for future work.

To conclude, the heat transfer between smoothly curved objects can be systematically expanded for small separations. This expansion is well suited for the gradient of the transfer, and a near-field adjusted plot allows accurate prediction of experimentally measurable data. If the separation is the smallest scale, the expansion can be performed analytically, where coefficients are evaluated numerically. 

We thank G.~Bimonte, T.~Emig, R.~L.~Jaffe, M.~F.~Maghrebi, M.~T.~H.~Reid and N.~Graham for helpful discussions. This research was supported by the NSF Grant No. DMR-12-06323, DOE grant No. DE-FG02-02ER45977, and the DFG grant No. KR 3844/1-1.

\bibliographystyle{apsrev}

\end{document}